\documentclass[aps,prl,twocolumn,superscriptaddress,showpacs]{revtex4} %

\newcommand{\figsize}{0.55\textwidth}      

\usepackage{graphicx}% Include figure files
\usepackage{dcolumn}% Align table columns on decimal point
\usepackage{bm}% bold math
\usepackage{lineno}
\usepackage{setspace}
\usepackage{amsfonts}
\usepackage{amssymb,amsmath}
\usepackage{xspace}
\usepackage{color}

\newcommand{\ppbar}  {\ensuremath{p\bar{p}}\xspace}
\newcommand{\qqbar}  {\ensuremath{q\bar{q}}\xspace}
\newcommand{\ttbar}  {\ensuremath{t\bar{t}}\xspace}

\newcommand{\sigmagg}  {\ensuremath{\sigma(gg \rightarrow \ttbar)}\xspace}
\newcommand{\sigmapp}  {\ensuremath{\sigma(\ppbar \rightarrow \ttbar)}\xspace}

\newcommand{\met}{\mbox{${\not\!\!E_{\rm T}}$}}

\newcommand{\cts}    {$\cos\theta^{*}$\xspace}

\newcommand{\Cs}    {\ensuremath{S_{f}}\xspace}

\newcommand{\barCs} {\ensuremath{\bar{\Cs}}\xspace}

\newcommand{\Cf}     {\ensuremath{G_{f}^{S}}\xspace}
\newcommand{\Cftrue} {\ensuremath{G_{f}}\xspace}

\newcommand{\Cffit}  {\ensuremath{G_{f}^{fit}}\xspace}

\newcommand{\Tgg} {\ensuremath{T^{gg}}\xspace}
\newcommand{\Tqq} {\ensuremath{T^{qq}}\xspace}
\newcommand{\TWJ} {\ensuremath{T^{bkg}}\xspace}

\newcommand{\unit}[1]{\ensuremath{\mathrm{\:#1}}}

\begin{document}

%%%%%%%%%%%%%%%%%%%%%%%%%%%%%%%%%%%%%%%%%%%%%%%
% Toggle double spacing and line numbering
% Won't work with the PRD revtex4 !
%\pagewiselinenumbers
%\doublespace
%%%%%%%%%%%%%%%%%%%%%%%%%%%%%%%%%%%%%%%%%%%%%%%

%uncomment for the PRD
\title{Measurement of the fraction of \ttbar production via gluon-gluon fusion in \ppbar collisions at $\sqrt{s}$ = 1.96\unit{TeV}}
\date{\today}
\affiliation{Institute of Physics, Academia Sinica, Taipei, Taiwan 11529, Republic of China} 
\affiliation{Argonne National Laboratory, Argonne, Illinois 60439} 
\affiliation{University of Athens, 157 71 Athens, Greece} 
\affiliation{Institut de Fisica d'Altes Energies, Universitat Autonoma de Barcelona, E-08193, Bellaterra (Barcelona), Spain} 
\affiliation{Baylor University, Waco, Texas  76798} 
\affiliation{Istituto Nazionale di Fisica Nucleare Bologna, $^t$University of Bologna, I-40127 Bologna, Italy} 
\affiliation{Brandeis University, Waltham, Massachusetts 02254} 
\affiliation{University of California, Davis, Davis, California  95616} 
\affiliation{University of California, Los Angeles, Los Angeles, California  90024} 
\affiliation{University of California, San Diego, La Jolla, California  92093} 
\affiliation{University of California, Santa Barbara, Santa Barbara, California 93106} 
\affiliation{Instituto de Fisica de Cantabria, CSIC-University of Cantabria, 39005 Santander, Spain} 
\affiliation{Carnegie Mellon University, Pittsburgh, PA  15213} 
\affiliation{Enrico Fermi Institute, University of Chicago, Chicago, Illinois 60637} 
\affiliation{Comenius University, 842 48 Bratislava, Slovakia; Institute of Experimental Physics, 040 01 Kosice, Slovakia} 
\affiliation{Joint Institute for Nuclear Research, RU-141980 Dubna, Russia} 
\affiliation{Duke University, Durham, North Carolina  27708} 
\affiliation{Fermi National Accelerator Laboratory, Batavia, Illinois 60510} 
\affiliation{University of Florida, Gainesville, Florida  32611} 
\affiliation{Laboratori Nazionali di Frascati, Istituto Nazionale di Fisica Nucleare, I-00044 Frascati, Italy} 
\affiliation{University of Geneva, CH-1211 Geneva 4, Switzerland} 
\affiliation{Glasgow University, Glasgow G12 8QQ, United Kingdom} 
\affiliation{Harvard University, Cambridge, Massachusetts 02138} 
\affiliation{Division of High Energy Physics, Department of Physics, University of Helsinki and Helsinki Institute of Physics, FIN-00014, Helsinki, Finland} 
\affiliation{University of Illinois, Urbana, Illinois 61801} 
\affiliation{The Johns Hopkins University, Baltimore, Maryland 21218} 
\affiliation{Institut f\"{u}r Experimentelle Kernphysik, Universit\"{a}t Karlsruhe, 76128 Karlsruhe, Germany} 
\affiliation{Center for High Energy Physics: Kyungpook National University, Daegu 702-701, Korea; Seoul National University, Seoul 151-742, Korea; Sungkyunkwan University, Suwon 440-746, Korea; Korea Institute of Science and Technology Information, Daejeon, 305-806, Korea; Chonnam National University, Gwangju, 500-757, Korea} 
\affiliation{Ernest Orlando Lawrence Berkeley National Laboratory, Berkeley, California 94720} 
\affiliation{University of Liverpool, Liverpool L69 7ZE, United Kingdom} 
\affiliation{University College London, London WC1E 6BT, United Kingdom} 
\affiliation{Centro de Investigaciones Energeticas Medioambientales y Tecnologicas, E-28040 Madrid, Spain} 
\affiliation{Massachusetts Institute of Technology, Cambridge, Massachusetts  02139} 
\affiliation{Institute of Particle Physics: McGill University, Montr\'{e}al, Canada H3A~2T8; and University of Toronto, Toronto, Canada M5S~1A7} 
\affiliation{University of Michigan, Ann Arbor, Michigan 48109} 
\affiliation{Michigan State University, East Lansing, Michigan  48824}
\affiliation{Institution for Theoretical and Experimental Physics, ITEP, Moscow 117259, Russia} 
\affiliation{University of New Mexico, Albuquerque, New Mexico 87131} 
\affiliation{Northwestern University, Evanston, Illinois  60208} 
\affiliation{The Ohio State University, Columbus, Ohio  43210} 
\affiliation{Okayama University, Okayama 700-8530, Japan} 
\affiliation{Osaka City University, Osaka 588, Japan} 
\affiliation{University of Oxford, Oxford OX1 3RH, United Kingdom} 
\affiliation{Istituto Nazionale di Fisica Nucleare, Sezione di Padova-Trento, $^u$University of Padova, I-35131 Padova, Italy} 
\affiliation{LPNHE, Universite Pierre et Marie Curie/IN2P3-CNRS, UMR7585, Paris, F-75252 France} 
\affiliation{University of Pennsylvania, Philadelphia, Pennsylvania 19104} 
\affiliation{Istituto Nazionale di Fisica Nucleare Pisa, $^q$University of Pisa, $^r$University of Siena and $^s$Scuola Normale Superiore, I-56127 Pisa, Italy} 
\affiliation{University of Pittsburgh, Pittsburgh, Pennsylvania 15260} 
\affiliation{Purdue University, West Lafayette, Indiana 47907} 
\affiliation{University of Rochester, Rochester, New York 14627} 
\affiliation{The Rockefeller University, New York, New York 10021} 

\affiliation{Istituto Nazionale di Fisica Nucleare, Sezione di Roma 1, $^v$Sapienza Universit\`{a} di Roma, I-00185 Roma, Italy} 

\affiliation{Rutgers University, Piscataway, New Jersey 08855} 
\affiliation{Texas A\&M University, College Station, Texas 77843} 
\affiliation{Istituto Nazionale di Fisica Nucleare Trieste/\ Udine, $^w$University of Trieste/\ Udine, Italy} 
\affiliation{University of Tsukuba, Tsukuba, Ibaraki 305, Japan} 
\affiliation{Tufts University, Medford, Massachusetts 02155} 
\affiliation{Waseda University, Tokyo 169, Japan} 
\affiliation{Wayne State University, Detroit, Michigan  48201} 
\affiliation{University of Wisconsin, Madison, Wisconsin 53706} 
\affiliation{Yale University, New Haven, Connecticut 06520} 
\author{T.~Aaltonen}
\affiliation{Division of High Energy Physics, Department of Physics, University of Helsinki and Helsinki Institute of Physics, FIN-00014, Helsinki, Finland}
\author{J.~Adelman}
\affiliation{Enrico Fermi Institute, University of Chicago, Chicago, Illinois 60637}
\author{T.~Akimoto}
\affiliation{University of Tsukuba, Tsukuba, Ibaraki 305, Japan}
\author{M.G.~Albrow}
\affiliation{Fermi National Accelerator Laboratory, Batavia, Illinois 60510}
\author{B.~\'{A}lvarez~Gonz\'{a}lez}
\affiliation{Instituto de Fisica de Cantabria, CSIC-University of Cantabria, 39005 Santander, Spain}
\author{S.~Amerio$^u$}
\affiliation{Istituto Nazionale di Fisica Nucleare, Sezione di Padova-Trento, $^u$University of Padova, I-35131 Padova, Italy} 

\author{D.~Amidei}
\affiliation{University of Michigan, Ann Arbor, Michigan 48109}
\author{A.~Anastassov}
\affiliation{Northwestern University, Evanston, Illinois  60208}
\author{A.~Annovi}
\affiliation{Laboratori Nazionali di Frascati, Istituto Nazionale di Fisica Nucleare, I-00044 Frascati, Italy}
\author{J.~Antos}
\affiliation{Comenius University, 842 48 Bratislava, Slovakia; Institute of Experimental Physics, 040 01 Kosice, Slovakia}
% ...

\collaboration{CDF Collaboration\footnote{With visitors from $^a$Universiteit Antwerpen, B-2610 Antwerp, Belgium, 
$^b$Chinese Academy of Sciences, Beijing 100864, China, 
$^c$University of Bristol, Bristol BS8 1TL, United Kingdom, 
$^d$University of California Irvine, Irvine, CA  92697, 
$^e$University of California Santa Cruz, Santa Cruz, CA  95064, 
$^f$Cornell University, Ithaca, NY  14853, 
$^g$University of Cyprus, Nicosia CY-1678, Cyprus, 
$^h$University College Dublin, Dublin 4, Ireland, 
$^i$University of Edinburgh, Edinburgh EH9 3JZ, United Kingdom, 
$^j$Universidad Iberoamericana, Mexico D.F., Mexico, 
$^k$University of Manchester, Manchester M13 9PL, England, 
$^l$Nagasaki Institute of Applied Science, Nagasaki, Japan, 
$^m$University de Oviedo, E-33007 Oviedo, Spain, 
$^n$Queen Mary, University of London, London, E1 4NS, England, 
$^o$Texas Tech University, Lubbock, TX  79409, 
$^p$IFIC(CSIC-Universitat de Valencia), 46071 Valencia, Spain,
$^x$Royal Society of Edinburgh/Scottish Executive Support Research Fellow,
$^y$Istituto Nazionale di Fisica Nucleare, Sezione di Cagliari, 09042 Monserrato (Cagliari), Italy 
}}
\noaffiliation

\begin{abstract}
We present a measurement of the ratio of the \ttbar production cross
section via gluon-gluon fusion to the total \ttbar production cross
section in \ppbar collisions at $\sqrt{s}$ = 1.96\unit{TeV} at the
Tevatron.  Using a data sample with an integrated luminosity of
955\unit{pb^{-1}} recorded by the CDF II detector at Fermilab, we
select events based on the \ttbar decay to lepton+jets.
Using an artificial neural network technique we discriminate between
\ttbar events produced via \qqbar annihilation and $gg$ fusion, and 
find \Cftrue=\sigmagg/\sigmapp $<0.33$ at the 68\% confidence level.
This result is combined with a previous measurement to obtain the most
stringent measurement of this quantity by CDF to date,
$\Cftrue=0.07^{+0.15}_{-0.07}$.

\end{abstract}

% taken from http://authors.aps.org/eprint/gateway/pacslist

%uncomment for the PRL
\maketitle

%%%%%%%%%%%%%%%%%%%%%%%%%%%%%%%%%%%%%%%%%%%%%%%%%%%%%%%%%%%
% \newpage
%              - Theory/previous results

In hadron colliders, such as the Tevatron, the pair production of heavy quarks has contributions from the different partons present in the initial state hadrons.  While for a given quark flavor the total pair production cross section can be measured simply by counting events in specific final-state channels, the contribution from the different primary partons to this cross section is difficult to estimate. 
For beauty, charm, and light-flavor production, the hadronization process does not normally allow the spin and kinematic properties of a quark to be observed through the analysis of the final state particles.  The situation is different for \ttbar production. The top quark, with a mass of about 175\unit{GeV/{\it c}^{2}}, has a lifetime that is an order of magnitude shorter than the typical hadronization time of $\approx 5 \times 10^{-24}\unit{s}$~\cite{c:PDG}. As a consequence, the spin and kinematic information of the top quark are preserved in its decay products, allowing the different \ttbar production processes to be distinguished based on the kinematic characteristics of the final state particles.

The standard model (SM) predicts the \ttbar production processes to be \qqbar annihilation ($\qqbar\rightarrow\ttbar$) and $gg$ fusion ($gg\rightarrow\ttbar$), occurring at the Tevatron with relative fractions of $\sim$85\% and $\sim$15\%, respectively, and having significantly different kinematic properties~\cite{c:MahlonParke}. 
Predictions for the relative fraction of \ttbar production from $gg$ fusion range from 10\% to 20\% due to uncertainties in the parton density functions~\cite{c:Mangano, c:Kidonakis}. 
A measurement of this fraction tests the SM prediction and our understanding of gluon parton distribution functions (PDF) in the proton. Disagreement with this prediction could reveal the possible existence of new mechanisms of top quark production and decay. For instance, production of top pairs at the Tevatron could be affected by a new vector particle associated with topcolor~\cite{c:BSM1,c:BSM2}. Such a resonance would affect the angular correlations between the top and antitop, and the relative mixture of \qqbar and $gg$ initiated \ttbar production.  Additionally, new physics in the decay of the top quark, such as a $t\rightarrow H^{+}~b$, would also affect these correlations ~\cite{c:BSM1}.

%This Letter details the first measurement of the fraction $\Cf=\sigmagg/\sigmapp$ based on the kinematics of the \ttbar decay products.  For the first time we use the kinematics of the production and decay of a \ttbar event to distinguish between the two modes of production.  This analysis is also described in detail in~\cite{c:Jared}.  A complementary measurement of this fraction described in~\cite{c:TrackAna} takes advantage of the higher probability for a primary gluon, compared to a quark, to radiate a low energy gluon in the production process.

This Letter details the first measurement of the fraction $\Cftrue=\sigmagg/\sigmapp$ based on the kinematics of \ttbar production, its decay products, and their correlations. We use the \ttbar event kinematics in an artificial neural network (NN) to distinguish between the two modes of production. This analysis is described in detail in~\cite{c:Jared}. 
We use data in $\ppbar$ collisions at $\sqrt{s}=1.96\unit{TeV}$ collected by the multi-purpose Collider Detector at Fermilab (CDF II)~\cite{c:CDFDet} from February 2002 to March 2006, corresponding to an integrated luminosity of 955\unit{pb^{-1}}.
The result of this analysis is then combined with a complementary measurement of this fraction~\cite{c:TrackAna}, which takes advantage of the higher probability for a primary gluon, compared to a quark, to radiate a low energy gluon in the production process.

%%%%%%%%%%%%%%%%%%%%%%%%%%%%%%%%%%%%%%%%%%%%%%%%%%%%%%%%%%%
%                     - Event selection
%    (ttbar signature and background contamination)
%We use data in $\ppbar$ collisions at $\sqrt{s}=1.96\unit{TeV}$ collected by the Collider Detector at Fermilab II (CDF II) from February 2002 to March 2006, corresponding to an integrated luminosity of 955\unit{pb^{-1}}. The CDF II detector is a multi-purpose particle detector and it is described in detail elsewhere~\cite{c:CDFDet}. 

% the actual event selection
%The production of \ttbar is followed almost always by the decay of each top quark to a $W$ boson and a $b$ quark~\cite{c:PDG}.  
The production of \ttbar is expected to be followed by the decay of each top quark to a $W$ boson and a $b$ quark with a branching ratio of approximately 100\% ~\cite{c:PDG}. We select events according to the topology of the \ttbar decay to lepton+jets, in which one $W$ decays leptonically and the other one hadronically, $\ttbar \rightarrow W^{+} b~ W^{-}\bar{b} \rightarrow l \nu_{l}~ b~ q'~ \bar{q}~ \bar{b}$. We require events to have an electron or muon candidate with  $p_{\rm T}>20\unit{GeV/{\it c}}$ and $|\eta|<1$, an imbalance of transverse energy of $\met>20\unit{GeV}$~\cite{c:coord} as expected from the undetected neutrino, and four or more jets with $p_{\rm T}>20\unit{GeV/{\it c}}$ and $|\eta|<2$ \cite{c:coord}.  A jet is defined as a cluster of energy in the calorimeter and is reconstructed using an algorithm with a fixed cone of radius 0.4 in $\eta$-$\phi$ space~\cite{c:jetclu}.  To account for non-linearities in the detector response and multiple $\ppbar$ collisions in an event, we correct jet energies and $\met$~\cite{c:jetcorr}.  Furthermore, to increase the purity of the sample, at least one jet in the event is required to have a displaced vertex ($b$-tag), which is indicative of the likely $b$-quark origin of the jet~\cite{c:Tags}.
% Events with 3 or more $b$-tags are ignored.
% Number of candidate events
We find 167 candidate events with one $b$-tag, and 65 candidate events with two or more $b$-tags that pass our event selection.

% Perhaps put more about how the sample composition was estimated
Background processes in the \ttbar candidate sample originate primarily from direct $W$+jets production, with minor contributions from diboson production ($WW$, $WZ$, $ZZ$), and multi-jet production (non-$W$). 
The expected background estimates in the one $b$-tag and two or more $b$-tags categories are shown in Table~\ref{t:XS} and were determined in~\cite{c:XS} with 318\unit{pb^{-1}} of data and scaled to 955\unit{pb^{-1}} of data.  We also show the number of events observed in the data and the signal fraction (\barCs). 
% Events with more than 2 $b$-tags are discarded.

\begin{table}
\begin{center}
\caption{Number of expected background events in the one $b$-tag and two or more $b$-tags categories for an integrated luminosity of 955\unit{pb^{-1}}. The number of events observed in the data and the signal fraction (\barCs) are also listed. \label{t:XS} }
%\TBC{The background numbers need to be revisited}
\begin{tabular}{c c c}
\hline
\hline
                     &  1 $b$-tag    &   $\geq2$ $b$-tags  \\
\hline
Total background     &  $27.3\pm3.4$   & $2.6\pm0.7$       \\
%\hline
%expected \ttbar      & $139\pm14$  & $65\pm20$ \\
%\hline
data     &      167       &     65    \\
%\hline
% the one below is the one considering Poisson fluctuations
%signal fraction \barCs & $0.84\pm0.11$  &   $0.96\pm0.42$  \\
signal fraction \barCs & $0.84\pm0.11$  &   $0.96\pm0.17$  \\
\hline
\hline
\end{tabular}
\end{center}
\end{table}

%%%%%%%%%%%%%%%%%%%%%%%%%%%%%%%%%%%%%%%%%%%%%%%%%%%%%%%%%%%
%     Monte Carlo datasets
Using the {\sc herwig} version 6.5~\cite{c:Herwig} leading-order (LO) Monte Carlo (MC) generator, we simulate \ttbar signal samples for the two production processes with a top quark mass of 175\unit{GeV/{\it c}^{2}}.  We model the dominant background of $W$+light and $W$+heavy flavor jets events with the {\sc alpgen} MC generator~\cite{c:Alpgen}, using {\sc herwig} to model parton showers.  All generated events are passed through the CDF II detector simulation~\cite{c:CDFDet}. 
%Table~\ref{t:XS} shows that the background
%consists predominantly of $W$+light and $W$+heavy flavor jets and the
%shapes of the non-$W$ and QCD jet backgrounds are very similar.
%Departure from this assumption is taken as a source of 
%systematic uncertainty.

%%%%%%%%%%%%%%%%%%%%%%%%%%%%%%%%%%%%%%%%%%%%%%%%%%%%%%%%%%%
%               - Method description/overview 

%%%%%%%%%%%%%%%%%%%%%%%%%%%%%%%%%%%%%%%%%%%%%%%%%%%%%%%%%%%
%       kinematic fitter
% The four-momenta of all the particles in the  \ttbar decay chain to lepton+jets can be reconstructed provided the final state particles are identified and measured.
 We consider only the four jets with highest transverse energy in each event and include all possible permutations associating jets with partons consistent with the $b$-tag information.
% The reconstruction does not depend on the association of jets to the decay products of the hadronically decaying $W$ boson. 
%There is also a two-fold ambiguity of the $p_{Z}$ component of the neutrino.  
The reconstruction is performed using a kinematic fitter~\cite{c:KinFitter} which compares the jet-to-parton association to the \ttbar hypothesis assuming the masses of the $W$ bosons and the top quarks to be 81\unit{GeV/{\it c}^{2}} and 175\unit{GeV/{\it c}^{2}}, respectively.  In the fitter the energy scale of the jets is varied according to its uncertainty. The agreement between each permutation and the \ttbar hypothesis is quantified by the $\chi^2$ value of the fit.  For each event the permutation with the lowest $\chi^2$ is used to extract kinematic variables as described below.

%%%%%%%%%%%%%%%%%%%%%%%%%%%%%%%%%%%%%%%%%%%%%%%%%%%%%%%%%%%
%     kinematic variables
We calculate eight variables that are sensitive to the production mechanism; two of these describe the production and the other six describe the decay of a given \ttbar event. 
%The production variables are evaluated in the \ttbar rest frame and are the cosine of the angle between the top quark momentum and the direction of the incoming proton, \cts, and the top quark velocity relative to $c$, $\beta$.  
At leading order the \ttbar production rate depends on two variables evaluated in the \ttbar rest frame; the cosine of the angle between the top quark momentum and the direction of the incoming proton, \cts, and the top quark velocity relative to $c$, $\beta$~\cite{c:offdiagonal2}. Since the functional form of the \ttbar production rate to these variables is different for $gg$ fusion than it is for \qqbar annihilation, these variables contain information that could allow us to distinguish the production mechanism.
The remaining six variables describe the \ttbar decay and contain information about the correlations between the spins of the top quarks.  These variables are the cosines of the angles with respect to the ``off-diagonal'' spin basis~\cite{c:offdiagonal2, c:offdiagonal} in the parent top-quark rest frame.  
%The ``off-diagonal'' spin basis is defined in the \ttbar rest frame. 
One characteristic feature of this basis is that the number of $\qqbar\rightarrow\ttbar$ events that have parallel top-quark spins as evaluated in this basis vanishes. Top pair events with parallel top-quark spins come exclusively from $gg$ production. The decay variables are the cosines of the angles between the direction of the ``off-diagonal'' basis and the lepton, neutrino, leptonically decaying $W$, down-type  quark, up-type quark, and hadronically decaying $W$. 
% : $\cos\theta_{Lep}$, $\cos\theta_{\nu}$, $\cos\theta_{Wlep}$, $\cos\theta_{Down}$, and $\cos\theta_{Up}$, and  $\cos\theta_{Whad}$.  
The distribution of data events for these variables shows very good agreement with the distributions from simulated background and \ttbar events with $\Cftrue=0.15$.

\begin{figure}
\includegraphics[width=\figsize]{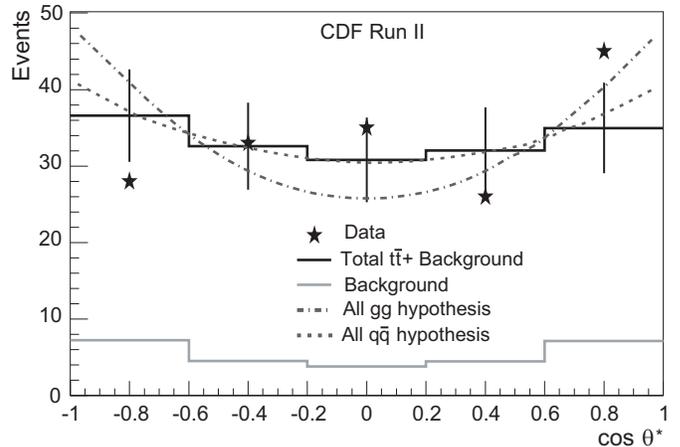}
\caption{Distributions of \cts for events with one $b$-tag in data, expected $\ttbar$ and background. 
The distribution of $\ttbar$ plus background uses the ratio of $gg$ to total \ttbar obtained from the fit of $\Cffit=-0.075$. We also show the expected distributions for $gg$-only and \qqbar-only hypotheses where background is included. The error bars on the total \ttbar + background includes the statistical uncertainty from Poisson statistics.}
\label{f:beta}
%\TBC{We should remake this plot so that it shows 100\% gg and 100\% qq (each normalized to expected number of ttbar events) and the bkg.}
\end{figure}

%%%%%%%%%%%%%%%%%%%%%%%%%%%%%%%%%%%%%%%%%%%%%%%%%%%%%%%%%%%
%     Neural Network
To obtain a single discriminating quantity the eight kinematic variables are fed into a NN.  The NN used in this analysis~\cite{c:ROOT} has an architecture of eight inputs, two hidden layers individually with ten and five nodes, and a single output. We train the NN to distinguish between \qqbar and $gg$ simulated \ttbar events, with separate training for the one $b$-tag events and for two or more $b$-tags events. 
Reducing or increasing the numbers of layers and nodes does not significantly change the discriminating power.  
Approximately one-third of the discriminating power comes from the production variable $\beta$, one-third from \cts, and one-third from the remaining six decay variables.  Figure~\ref{f:beta} shows the distribution of \cts, one of our more sensitive variables, for events with one $b$-tag in data, expected $\ttbar$, and background.
%Approximately 36\% of the discriminating power comes from the production variable $\beta$, 29\% from \cts, and 35\% from the remaining six decay variables.  Figure~\ref{f:beta} shows the distribution of \cts, one of our more sensitive variables, for events with one $b$-tag in data, expected $\ttbar$, and background.

%%%%%%%%%%%%%%%%%%%%%%%%%%%%%%%%%%%%%%%%%%%%%%%%%%%%%%%%%%%
%     NN Templates
In each $b$-tag category we obtain three template distributions of the NN output, \Tqq, \Tgg, and \TWJ,  by running the NN over \qqbar produced \ttbar, $gg$ produced \ttbar, and background MC events, respectively.  These templates represent the probability for an event to have the NN output obtained assuming it was a \ttbar produced by $gg$ fusion, a \ttbar produced by $\qqbar$ annihilation, or a background event.

%%%%%%%%%%%%%%%%%%%%%%%%%%%%%%%%%%%%%%%%%%%%%%%%%%%%%%%%%%%
%     Likelihood
An estimator of the $gg$ fraction in the sample, \Cf,  is obtained by maximizing a likelihood function.  We calculate the likelihood of the full event sample for a given \Cf as 
\begin{equation}
\mathcal{L}(\Cf) = \mathcal{L}^{1}(\Cf) \mathcal{L}^{2}(\Cf),
\label{e:KinLike}
\end{equation}
where $\mathcal{L}^{1}$ and $\mathcal{L}^{2}$ are the one $b$-tag and two or more $b$-tags likelihoods, respectively. The individual likelihoods are defined as
\begin{eqnarray}
%\mathcal{L}^{\mathcal{N}b}(\Cf) & = & e^{\frac{-(\Cs^{\mathcal{N}b}-\barCs^{\mathcal{N}b})^2}{2*\sigma_{\Cs^{\mathcal{N}b}}^2}} \prod \left\{ (1- \Cs^{\mathcal{N}b}) \TWJ_{\mathcal{N}b} \right. \nonumber \\
% & + & \left. \Cs^{\mathcal{N}b}  \left[ \Cf \Tgg_{\mathcal{N}b}  +(1-\Cf) \Tqq_{\mathcal{N}b} \right]  \right\},
\mathcal{L}^{i}(\Cf) & = & \exp\left[\frac{-(\Cs^{i}-\barCs^{i})^2}{2 \sigma_{\barCs^{i}}^2}\right] \prod \left\{ (1- \Cs^{i}) \TWJ_{i} \right. \nonumber \\
& + & \left. \Cs^{i}  \left[ \Cf \Tgg_{i}  +(1-\Cf) \Tqq_{i} \right]  \right\},
\end{eqnarray}
where $i=\{1,2\}$, the product is over the events in the $i$th $b$-tag category, the values of the signal fraction $\barCs^{i}$ and its uncertainty $\sigma_{\barCs^{i}}^{2}$ are taken from Table~\ref{t:XS}, and the variables $\Cs^{i}$ represent the observed signal fractions in the sample.
% ORIGINAL where $i=\{1,2\}$, the product is over the events in the $i$th $b$-tag category, and the values of the signal fraction $\barCs^{i}$ and its error $\sigma_{\Cs^{i}}^{2}$ are taken from Table~\ref{t:XS}.
The overall multiplicative term represents a Gaussian weight centered at the expected signal fraction.  
Scanning over values of \Cf we find the maximum likelihood solution by varying the fractions $\Cs^{1}$ and $\Cs^{2}$. In a given sample the \Cf value for which the likelihood is maximum is called \Cffit.
%%%%%%%%%%%%%%%%%%%%%%%%%%%%%%%%%%%%%%%%%%%%%%%%%%%%%%%%%%%
%       Feldman-Cousins and PSE's
The fitted fraction \Cffit is related to the true production fraction \Cftrue by the acceptance ratio of $gg\rightarrow\ttbar$ to $\qqbar\rightarrow\ttbar$. Using {\sc herwig} MC \cite{c:Herwig} calculations we estimate the acceptance ratio to be $1.29\pm0.02$ and  $1.25\pm0.02$ for the one $b$-tag and two or more $b$-tags categories respectively.
The value of \Cffit is not constrained to the physically allowable range between zero and unity, and neither would be an estimate of \Cftrue obtained from taking into account the acceptance ratio.
To ensure a result for \Cftrue in the physical range we use the Feldman-Cousins prescription (FC)~\cite{c:FC}, which maps any result of \Cffit to a range of the true fraction \Cftrue. We generate this map by fitting for \Cffit in simulated experiments with a known  \Cftrue and $\barCs^{i}$. These simulated experiments are a mixture of $gg$, \qqbar, and background events keeping the total number in each experiment fixed to that observed in data for that $b$-tag category.  We fit the distribution of \Cffit for the simulated experiments to a Gaussian shape for each value of \Cftrue.  The FC likelihood-ratio ordering principle~\cite{c:FC} is applied to the Gaussian obtained from the simulated experiments to construct the confidence level (C.L.) bands.
%%%%%%%%%%%%%%%%%%%%%%%%%%%%%%%%%%%%%%%%%%%%%%%%%%%%%%%%%%%
%     RESULTS
%The Feldman-Cousins likelihood-ratio ordering principle~\cite{c:FC} is applied to the Gaussian obtained from the simualated experiments to construct the confidence level (C.L.) bands shown in Fig.~\ref{f:Final_FC}.  
%These bands, represent the statistical uncertainty alone and also the combined statistical and systematic uncertainties at difference confidence levels.  The minimum of the negative log likelihood for the data sample is $\Cffit=-0.075$.  For this value of \Cffit the FC construction shown in Fig.~\ref{f:Final_FC} results in values of $\Cftrue<0.33$ at the 68\% C.L. and $\Cftrue<0.61$ at the 95\% C.L., respectively.

%%%%%%%%%%%%%%%%%%%%%%%%%%%%%%%%%%%%%%%%%%%%%%%%%%%%%%%%%%%
%     SYSTEMATIC UNCERTAINTIES
To incorporate systematic effects into the FC prescription we generate auxiliary sets of simulated experiments chosen from signal and/or background samples designed to study various sources of systematic uncertainty. The difference of the mean of the \Cffit distributions generated with the standard and the auxiliary simulated experiments is added in quadrature to the original width of the Gaussian distribution obtained with the standard sample.  We repeat the procedure for each value of \Cftrue and for significant sources of systematic uncertainties. 
 The dominant systematic uncertainties result from uncertainties in the background shape and composition, and differences between LO and next-to-leading order (NLO) predictions estimated by comparing to a \ttbar MC sample generated with the NLO generator {\sc mc@nlo}~\cite{c:mcnlo}. We evaluate the systematic uncertainty on this measurement due to parton distribution functions by using MC samples generated with MRST PDF's~\cite{c:mrst} and the full set  of eigenvectors known as CTEQ6M from the CTEQ collaboration~\cite{c:cteq}.  We also include sources of systematic effects arising from the jet energy scale~\cite{c:jetcorr} and initial and final state radiation~\cite{c:isrfsr}.

%%%%%%%%%%%%%%%%%%%%%%%%%%%%%%%%%%%%%%%%%%%%%%%%%%%%%%%%%%%
%     RESULTS
Finally, we evaluate the log likelihood for the data sample and find the minimum of the negative log likelihood to be $\Cffit=-0.075$. 
The variables $\Cs^{1}$ and $\Cs^{2}$ for this value of $\Cffit$ match those given in 
Table~\ref{t:XS} within the uncertainties. The $\chi^2$  goodness of fit test between the observed data values and the expected values at \Cffit results in $\chi^2/ndf=0.9$, indicating a good agreement between the observed and fitted distributions. For this value of \Cffit the FC construction results in $\Cftrue<0.33$ at the 68\% C.L. and $\Cftrue<0.61$ at the 95\% C.L., respectively.

\begin{figure}
\includegraphics[width=\figsize]{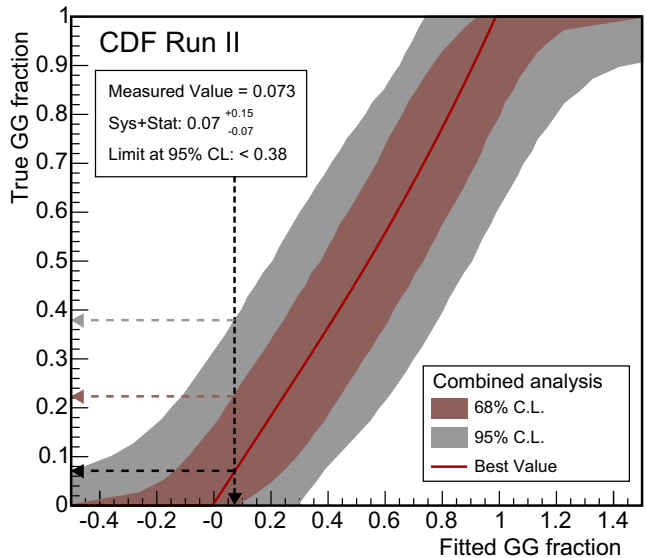}
\caption{Feldman-Cousins bands for the combination of the analyses with statistical and systematic uncertainties for 68\% C.L. and
95\% C.L.. In the data, we find $\Cffit=0.073$, which yields $\Cftrue=0.07^{+0.15}_{-0.07}$ and $\Cftrue<0.38$ at the 95\% C.L.}
\label{f:Final_FC}
\end{figure}

%%%%%%%%%%%%%%%%%%%%%%%%%%%%%%%%%%%%%%%%%%%%%%%%%%%%%%%%%%%
%     Combination
This measurement is combined with the one performed
in~\cite{c:TrackAna} also using 955\unit{pb^{-1}} of CDF data, in
which the \Cffit fraction is estimated from a fit to the distribution
of the number of low transverse momentum charged particles in the
event by comparing the data distribution to those from
gluon-originated and quark-originated processes.  This analysis alone
results in $\Cffit=0.09\pm0.16$, where the uncertainty includes the statistical and 
systematic uncertainties.

We perform the combination using the FC prescription by including the
track multiplicity information in the simulated experiments used for the
evaluation of the FC bands.  The statistical correlation between the
event kinematics analysis and the track multiplicity analysis is found
to be negligible. For each $gg$ (\qqbar) produced
\ttbar event in a simulated experiment, the value of the track multiplicity for that event is
chosen randomly from the gluon-(quark-)originated track
distribution. Primary gluons are estimated to contribute to background
processes in $54\pm9$\% of the cases~\cite{c:TrackAna}. Therefore,
for background events in a given simulated experiment the value of the
track multiplicity is obtained from the gluon-originated distribution
54\% of the time and from the quark distribution the remaining times.
For each simulated experiment we evaluate the likelihood as a function of
\Cf for the track multiplicity analysis using the goodness-of-fit to
data for that fraction.  We construct the combined likelihood by
multiplying the likelihood of Eqn.~\ref{e:KinLike} by the
corresponding likelihood for the track multiplicity analysis.  The
distribution of \Cffit values of the combined likelihood is then used
to construct the combined FC bands shown in Fig.~\ref{f:Final_FC} at
68\% and 95\% C.L.  The value that maximizes the combined likelihood
for the observed events is $\Cffit=0.073$, indicated by the vertical
arrow in Fig.~\ref{f:Final_FC}.  For this value of \Cffit we measure
$\Cftrue=0.07^{+0.15}_{-0.07}$, and we find the 95\% C.L. limit to be
$\Cftrue<0.38$~\cite{c:StatTreat}.

%%%%%%%%%%%%%%%%%%%%%%%%%%%%%%%%%%%%%%%%%%%%%%%%%%%%%%%%%%%
%     Final comments
%\TBC{Put the final conclusions here!}

To conclude, we report on the first limit of the fraction of $gg$
produced \ttbar events relative to the total by differentiating
between the kinematic properties and their correlations of both
production processes. Using this technique we limit the fraction
$\Cftrue<0.33$ at 68\% C.L. and find it to be consistent with SM
expectations. The combination with the measurement described
in~\cite{c:TrackAna} results in $\Cftrue=0.07^{+0.15}_{-0.07}$,
yielding the most stringent measurement by CDF of this quantity to
date.

%%%%%%%%%%%%%%%%%%%%%%%%%%%%%%%%%%%%%%%%%%%%%%%%%%%%%%%%%%%
%     Acknowledgements
%check for the latest@
%http://www-cdf.fnal.gov/internal/people/links/CarolPicciolo/standardack.html

We thank the Fermilab staff and the technical staffs of the participating institutions for their vital contributions. This work was supported by the U.S. Department of Energy and National Science Foundation; the Italian Istituto Nazionale di Fisica Nucleare; the Ministry of Education, Culture, Sports, Science and Technology of Japan; the Natural Sciences and Engineering Research Council of Canada; the National Science Council of the Republic of China; the Swiss National Science Foundation; the A.P. Sloan Foundation; the Bundesministerium f\"ur Bildung und Forschung, Germany; the Korean Science and Engineering Foundation and the Korean Research Foundation; the Science and Technology Facilities Council and the Royal Society, UK; the Institut National de Physique Nucleaire et Physique des Particules/CNRS; the Russian Foundation for Basic Research; the Ministerio de Educaci\'{o}n y Ciencia and Programa Consolider-Ingenio 2010, Spain; the Slovak R\&D Agency; and the Academy of Finland.

%\end{acknowledgments}

%%%%%%%%%%%%%%%%%%%%%%%%%%%%%%%%%%%%%%%%%%%%%%%%%%%%%%%%%
%     References
\bibliography{GGVsQQPRL}

\begin{thebibliography}{26}
\expandafter\ifx\csname natexlab\endcsname\relax\def\natexlab#1{#1}\fi
\expandafter\ifx\csname bibnamefont\endcsname\relax
  \def\bibnamefont#1{#1}\fi
\expandafter\ifx\csname bibfnamefont\endcsname\relax
  \def\bibfnamefont#1{#1}\fi
\expandafter\ifx\csname citenamefont\endcsname\relax
  \def\citenamefont#1{#1}\fi
\expandafter\ifx\csname url\endcsname\relax
  \def\url#1{\texttt{#1}}\fi
\expandafter\ifx\csname urlprefix\endcsname\relax\def\urlprefix{URL }\fi
\providecommand{\bibinfo}[2]{#2}
\providecommand{\eprint}[2][]{\url{#2}}

\bibitem[{\citenamefont{Yao}(2006)}]{c:PDG}
\bibinfo{author}{\bibfnamefont{Y.-M.} \bibnamefont{Yao}}, \bibinfo{journal}{J.
  Phys. G.} \textbf{\bibinfo{volume}{33}}, \bibinfo{pages}{1232}
  (\bibinfo{year}{2006}).

\bibitem[{\citenamefont{Mahlon and Parke}(1996)}]{c:MahlonParke}
\bibinfo{author}{\bibfnamefont{G.}~\bibnamefont{Mahlon}} \bibnamefont{and}
  \bibinfo{author}{\bibfnamefont{S.}~\bibnamefont{Parke}},
  \bibinfo{journal}{Phys. Rev. D} \textbf{\bibinfo{volume}{53}},
  \bibinfo{pages}{4886} (\bibinfo{year}{1996}).

\bibitem[{\citenamefont{Cacciari~{\it et al.}}(2004)}]{c:Mangano}
\bibinfo{author}{\bibfnamefont{M.}~\bibnamefont{Cacciari~{\it et al.}}},
  \bibinfo{journal}{J. High Energy Phys.} \textbf{\bibinfo{volume}{0404}},
  \bibinfo{pages}{068} (\bibinfo{year}{2004}).

\bibitem[{\citenamefont{Kidonakis and Vogt}(2003)}]{c:Kidonakis}
\bibinfo{author}{\bibfnamefont{N.}~\bibnamefont{Kidonakis}} \bibnamefont{and}
  \bibinfo{author}{\bibfnamefont{R.}~\bibnamefont{Vogt}},
  \bibinfo{journal}{Phys. Rev. D} \textbf{\bibinfo{volume}{68}},
  \bibinfo{pages}{114014} (\bibinfo{year}{2003}).

\bibitem[{\citenamefont{Hill and Parke}(1994)}]{c:BSM1}
\bibinfo{author}{\bibfnamefont{C.~T.} \bibnamefont{Hill}} \bibnamefont{and}
  \bibinfo{author}{\bibfnamefont{S.~J.} \bibnamefont{Parke}},
  \bibinfo{journal}{Phys. Rev. D} \textbf{\bibinfo{volume}{49}},
  \bibinfo{pages}{4454} (\bibinfo{year}{1994}).

\bibitem[{\citenamefont{Lane and Eichten}(1995)}]{c:BSM2}
\bibinfo{author}{\bibfnamefont{K.~D.} \bibnamefont{Lane}} \bibnamefont{and}
  \bibinfo{author}{\bibfnamefont{E.}~\bibnamefont{Eichten}},
  \bibinfo{journal}{Phys. Lett.} \textbf{\bibinfo{volume}{B352}},
  \bibinfo{pages}{382} (\bibinfo{year}{1995}).

\bibitem[{\citenamefont{Yamaoka}(2007)}]{c:Jared}
\bibinfo{author}{\bibfnamefont{J.~A.} \bibnamefont{Yamaoka}}, Ph.D. thesis,
  \bibinfo{school}{Rutgers University} (\bibinfo{year}{2007}),
  \eprint{FERMILAB-THESIS-2007-18}.

\bibitem[{\citenamefont{Acosta et~al.}(2005)}]{c:CDFDet}
\bibinfo{author}{\bibfnamefont{D.~E.} \bibnamefont{Acosta}}
  \bibnamefont{et~al.} (\bibinfo{collaboration}{CDF}), \bibinfo{journal}{Phys.
  Rev. D} \textbf{\bibinfo{volume}{71}}, \bibinfo{pages}{052003}
  (\bibinfo{year}{2005}), \eprint{hep-ex/0410041}.

\bibitem[{\citenamefont{Aaltonen~{\it et al.}}(2008)}]{c:TrackAna}
\bibinfo{author}{\bibfnamefont{T.}~\bibnamefont{Aaltonen~{\it et al.}}}
  (\bibinfo{collaboration}{CDF Collaboration}), \bibinfo{journal}{Phys. Rev. D}
  \textbf{\bibinfo{volume}{78}}, \bibinfo{pages}{111101}
  (\bibinfo{year}{2008}).

\bibitem[{c:c()}]{c:coord}
\bibinfo{note}{CDF uses a ($z$, $\phi$, $\theta$) coordinate system with the
  $z$-axis in the direction of the proton beam; $\phi$ and $\theta$ are the
  azimuthal and polar angle respectively. The pseudorapidity is defined as
  $\eta = -\ln(\tan{\theta \over 2})$, and the transverse momentum and energy
  as $p_{\rm T} = p \sin \theta$ and $E_{\rm T} = E \sin \theta$, respectively.
  Missing transverse energy ($\met = |\metvec|$) is defined as $\metvec = -
  \sum_{i} E_{\rm T}^i {\bf{\hat{n}_i}}$ where ${\bf{\hat{n}_i}}$ is a unit
  vector in the transverse plane that points from the beam-line to the $i^{th}$
  calorimeter tower.}

\bibitem[{\citenamefont{Abe~{\it et al.}}(1992)}]{c:jetclu}
\bibinfo{author}{\bibfnamefont{F.}~\bibnamefont{Abe~{\it et al.}}}
  (\bibinfo{collaboration}{CDF Collaboration}), \bibinfo{journal}{Phys. Rev. D}
  \textbf{\bibinfo{volume}{45}}, \bibinfo{pages}{1448} (\bibinfo{year}{1992}).

\bibitem[{\citenamefont{Bhatti~{\it et al.}}(2006)}]{c:jetcorr}
\bibinfo{author}{\bibfnamefont{A.}~\bibnamefont{Bhatti~{\it et al.}}},
  \bibinfo{journal}{Nucl. Instrum. Methods A} \textbf{\bibinfo{volume}{566}},
  \bibinfo{pages}{375} (\bibinfo{year}{2006}).

\bibitem[{c:T()}]{c:Tags}
\bibinfo{note}{A jet is determined to be tagged if it shows a displaced
  secondary vertex. These jets typically originate from the decay of long lived
  mesons such as those resulting after the hadronization process of the $b$
  quark.}

\bibitem[{\citenamefont{Abulencia~{\it et al.}}(2006{\natexlab{a}})}]{c:XS}
\bibinfo{author}{\bibfnamefont{A.}~\bibnamefont{Abulencia~{\it et al.}}}
  (\bibinfo{collaboration}{CDF Collaboration}), \bibinfo{journal}{Phys. Rev.
  Lett.} \textbf{\bibinfo{volume}{97}}, \bibinfo{pages}{082004}
  (\bibinfo{year}{2006}{\natexlab{a}}).

\bibitem[{\citenamefont{Corcella~{\it et al.}}(2001)}]{c:Herwig}
\bibinfo{author}{\bibfnamefont{G.}~\bibnamefont{Corcella~{\it et al.}}},
  \bibinfo{journal}{J. High Energy Phys.} \textbf{\bibinfo{volume}{0101}},
  \bibinfo{pages}{010} (\bibinfo{year}{2001}).

\bibitem[{\citenamefont{Mangano~{\it et al.}}(2003)}]{c:Alpgen}
\bibinfo{author}{\bibfnamefont{M.~L.} \bibnamefont{Mangano~{\it et al.}}},
  \bibinfo{journal}{J. High Energy Phys.} \textbf{\bibinfo{volume}{0307}},
  \bibinfo{pages}{001} (\bibinfo{year}{2003}).

\bibitem[{\citenamefont{Abulencia~{\it et
  al.}}(2006{\natexlab{b}})}]{c:KinFitter}
\bibinfo{author}{\bibfnamefont{A.}~\bibnamefont{Abulencia~{\it et al.}}}
  (\bibinfo{collaboration}{CDF Collaboration}), \bibinfo{journal}{Phys. Rev. D}
  \textbf{\bibinfo{volume}{73}}, \bibinfo{pages}{032003}
  (\bibinfo{year}{2006}{\natexlab{b}}).

\bibitem[{\citenamefont{Mahlon and Parke}(1997)}]{c:offdiagonal2}
\bibinfo{author}{\bibfnamefont{G.}~\bibnamefont{Mahlon}} \bibnamefont{and}
  \bibinfo{author}{\bibfnamefont{S.}~\bibnamefont{Parke}},
  \bibinfo{journal}{Phys. Lett. B} \textbf{\bibinfo{volume}{411}},
  \bibinfo{pages}{173} (\bibinfo{year}{1997}).

\bibitem[{\citenamefont{Parke and Shadmi}(1996)}]{c:offdiagonal}
\bibinfo{author}{\bibfnamefont{S.}~\bibnamefont{Parke}} \bibnamefont{and}
  \bibinfo{author}{\bibfnamefont{Y.}~\bibnamefont{Shadmi}},
  \bibinfo{journal}{Phys. Lett. B} \textbf{\bibinfo{volume}{387}},
  \bibinfo{pages}{199} (\bibinfo{year}{1996}).

\bibitem[{\citenamefont{Brun and Rademakers}(1997)}]{c:ROOT}
\bibinfo{author}{\bibfnamefont{R.}~\bibnamefont{Brun}} \bibnamefont{and}
  \bibinfo{author}{\bibfnamefont{F.}~\bibnamefont{Rademakers}},
  \bibinfo{journal}{Nucl. Instrum. Methods A} \textbf{\bibinfo{volume}{389}},
  \bibinfo{pages}{81} (\bibinfo{year}{1997}), \bibinfo{note}{[The NN in this
  analysis is implemented using the {\sc TMultiLayerPerceptron} class.]}.

\bibitem[{\citenamefont{Feldman and Cousins}(1998)}]{c:FC}
\bibinfo{author}{\bibfnamefont{G.~J.} \bibnamefont{Feldman}} \bibnamefont{and}
  \bibinfo{author}{\bibfnamefont{R.~D.} \bibnamefont{Cousins}},
  \bibinfo{journal}{Phys. Rev. D} \textbf{\bibinfo{volume}{57}},
  \bibinfo{pages}{3873} (\bibinfo{year}{1998}).

\bibitem[{\citenamefont{Frixione et~al.}(2003)\citenamefont{Frixione, Nason,
  and Webber}}]{c:mcnlo}
\bibinfo{author}{\bibfnamefont{S.}~\bibnamefont{Frixione}},
  \bibinfo{author}{\bibfnamefont{P.}~\bibnamefont{Nason}}, \bibnamefont{and}
  \bibinfo{author}{\bibfnamefont{B.~R.} \bibnamefont{Webber}},
  \bibinfo{journal}{J. High Energy Phys.} \textbf{\bibinfo{volume}{08}},
  \bibinfo{pages}{007} (\bibinfo{year}{2003}).

\bibitem[{\citenamefont{Martin et~al.}(2000)\citenamefont{Martin, Roberts,
  Stirling, and Thorne}}]{c:mrst}
\bibinfo{author}{\bibfnamefont{A.~D.} \bibnamefont{Martin}},
  \bibinfo{author}{\bibfnamefont{R.~G.} \bibnamefont{Roberts}},
  \bibinfo{author}{\bibfnamefont{W.~J.} \bibnamefont{Stirling}},
  \bibnamefont{and} \bibinfo{author}{\bibfnamefont{R.~S.}
  \bibnamefont{Thorne}}, \bibinfo{journal}{Eur. Phys. J. C}
  \textbf{\bibinfo{volume}{14}}, \bibinfo{pages}{133} (\bibinfo{year}{2000}).

\bibitem[{\citenamefont{Pumplin~{\it et al.}}(2002)}]{c:cteq}
\bibinfo{author}{\bibfnamefont{J.}~\bibnamefont{Pumplin~{\it et al.}}},
  \bibinfo{journal}{Nucl. Intrum. Methods A} \textbf{\bibinfo{volume}{447}},
  \bibinfo{pages}{1} (\bibinfo{year}{2002}).

\bibitem[{\citenamefont{Abulencia~{\it et al.}}(2006{\natexlab{c}})}]{c:isrfsr}
\bibinfo{author}{\bibfnamefont{A.}~\bibnamefont{Abulencia~{\it et al.}}}
  (\bibinfo{collaboration}{CDF Collaboration}), \bibinfo{journal}{Phys. Rev. D}
  \textbf{\bibinfo{volume}{73}}, \bibinfo{pages}{032003}
  (\bibinfo{year}{2006}{\natexlab{c}}).

\bibitem[{c:S()}]{c:StatTreat}
\bibinfo{note}{The result presented in \cite{c:TrackAna}, which uses a Bayesian
  method with a prior that includes the physical boundaries, finds the 95\% CL
  to be $G_{f}<0.33$. In this Letter we use the Feldman-Cousins approach
  finding a combined 95\% CL of $G_{f}<0.38$, where the seemingly lower
  discriminating power stems only from the difference in the statistical
  treatment of both analyses. The result presented in \cite{c:TrackAna}
  evaluated with the Feldman-Cousins approach yields a 95\% CL of
  $G_{f}<0.42$.}

\end{thebibliography}

\end{document}